\documentclass[twocolumn,pra,showpacs]{revtex4}

\usepackage{epsfig}

\usepackage{graphicx}
\usepackage{amsmath}
\usepackage{amssymb}
\usepackage{psfrag}
\usepackage[sans]{dsfont}

\newcommand{\bra    }{\langle}
\newcommand{\ket    }{\rangle}
\newcommand{\idty}{\b 1}

\DeclareMathOperator{\Entropy}{S}
\newcommand{\dop}[1]{\ensuremath{|#1\hspace{-0.2ex}\rangle\hspace{-0.6ex}\langle\hspace{-0.1ex}#1|}}
\newcommand{\unity}{\mathds{1}}
\newcommand{\eref}[1]{Eq. (\ref{#1})}
\newcommand{\secref}[1]{(Sec. \ref{#1}, Pg. \pageref{#1})}

\newcommand{\comment}[1]{}

\begin{document}

\title{Classical capacity of a qubit depolarizing channel with memory}
\date{\today}

\author{Jeroen Wouters}\email{jeroen.wouters@fys.kuleuven.be}
\author{Mark Fannes}\email{mark.fannes@fys.kuleuven.be}
\affiliation{Instituut voor Theoretische Fysica, Katholieke Universiteit Leuven, Celestijnenlaan 200D, B-3001 Heverlee, Belgium}

\author{Ismail Akhalwaya}\email{akhalwaya@ukzn.ac.za}
\affiliation{Quantum Research Group, School of Physics, University of KwaZulu-Natal, Private Bag X54001, Durban, 4000, RSA; and \\Centre for High Performance Computing, 15 Lower Hope Rd, Rosebank, Cape Town, 7700, RSA}
\author{Francesco Petruccione}\email{petruccione@ukzn.ac.za}
\affiliation{Quantum Research Group and National Institute for Theoretical Physics, School of Physics, University of KwaZulu-Natal, Private Bag X54001, Durban, 4000, RSA}

\begin{abstract}
The classical product state capacity of a noisy quantum channel with memory is investigated. A forgetful noise-memory channel is constructed by Markov switching between two depolarizing channels which introduces non-Markovian noise correlations between successive channel uses. The computation of the capacity is reduced to an entropy computation for a function of a Markov process. A reformulation in terms of algebraic measures then enables its calculation. The effects of the hidden-Markovian memory on the capacity are explored. An increase in noise-correlations is found to increase the capacity.
\end{abstract}


\maketitle

\section{Introduction}

Quantum mechanics brings strange and wonderful features to the field of information theory. It introduces new information resources such as qubits with the power of superposition but also teasing restrictions such as the no-cloning theorem. We are interested in the possibility of the boosted transmission of classical information through a quantum channel with memory and no prior entanglement.

Great strides have been made in understanding the capacity of quantum channels. For example, the celebrated Holevo-Schumacher-Westmoreland (HSW) theorem \cite{HSW} gives an expression for the classical capacity of a noisy memoryless quantum channel with product state inputs. The memoryless channel restriction has since been extended to, so called, \textit{forgetful} memory channels \cite{DKRW}. The inclusion of memory is the next step in the attempt of accurately modelling the complicated noise-correlated real world. Now that these initial seeds of the theoretical framework are in place, it is enlightening to use these tools, in specific cases, to analytically study the new effects that noise with memory has on the capacity.

We construct a forgetful channel and incorporate memory effects by Markov switching between two sub-channels. In order to investigate the classical product state capacity of this channel we must look at the entropy of the classical output. The output sequence of qubits and their associated errors are correlated. To manage this complicated conditional dependence, we use the hidden Markov nature of the process to reformulate the problem using the algebraic measure construction \cite{FNS91}. The algebraic measure approach allows us to derive an expression for the asymptotic entropy rate. We then explore the effects that our non-Markovian memory has on the classical product state capacity.

This paper is structured as follows. In Section \ref{sec:classical_capacity}, we take a closer look at the quantity we are investigating, namely the product state classical capacity. In Section \ref{sec:depolarizing}, we construct the forgetful channel with Markovian noise correlations. In Section \ref{sec:algebraic}, algebraic measures are introduced, which are used in Section \ref{sec:our_algebraic} to reformulate the problem. Finally, in Section \ref{sec:results}, we show how this allows us to easily calculate the capacity of the channel numerically.

\section{Classical Capacity of Quantum Channels}
\label{sec:classical_capacity}
The information process we are studying is classical communication through a noisy quantum channel. The layout of this section largely follows that in \cite{HSW}.

With the classical information we want to send encoded using an input alphabet $A=\lbrace 1, \ldots, a \rbrace$, we choose for every element $i \in A$ an encoding quantum state $\rho_i$ on a Hilbert space $\mathfrak{H}$. This input state is then transmitted using a quantum channel $\Lambda: \mathcal{B}(\mathfrak{H}) \rightarrow  \mathcal{B}(\mathfrak{K})$. For the channel to be a valid quantum channel it must be a completely positive trace preserving map.

Transmitting the element $i \in A$ results in a quantum state $R_i=\Lambda(\rho_i)$ being received on the output side. On this side, the received quantum state is measured using a resolution of identity in $\mathfrak{K}$. This resolution of identity is a set of positive operators $X = \lbrace X_i \rbrace$ on $\mathfrak{K}$ such that $\sum_i X_i = \unity$.

The conditional probability of the receiver measuring $j$, when the input $i$ was sent, is given by $p(j|i) = \text{Tr}\, R_i X_j$. If at the input side the element $i$ is sent with a probability $\pi_i$, the amount of information that will be received is quantified by the classical Shannon information,
\begin{equation}
\label{eq:single_shannon}
I_{\Lambda,1}(\pi,\rho,X)= \sum_{i,j \in A} \pi_i p(i|j) \log \left( \frac{p(j|i)}{\sum_{k\in A} \pi_k p(j|k)} \right) \;.
\end{equation}

If the sender is allowed to use the channel $n$ times, the channel use can be described by the product channel $\Lambda_n = \otimes^n \Lambda$ on $\otimes^n \mathfrak{H}= \mathfrak{H} \otimes \ldots \otimes \mathfrak{H}$. The input alphabet is now $A^n$ and the probability distribution of a word $u=(i_1,\ldots,i_n)\in A^n$ being sent is again denoted by $\pi_u$. The codeword corresponding to the input $u$ is given by
\[ \rho_u= \rho_{i_1} \otimes \ldots \otimes \rho_{i_n}\]
and results in $ R_u = R_{i_1} \otimes \ldots \otimes R_{i_n}$ being received. The conditional probability and the Shannon information $I_{\Lambda,n}$ for the $n$-product of the channel can now be introduced completely analogously to \eref{eq:single_shannon}, with the summations over $A^n$ instead of $A$.

The maximum amount of information that can be sent with $n$ channel uses is now given by
\[ C_n(\Lambda) = \sup_{\pi,\rho,X} I_{\Lambda,n}(\pi,\rho,X) \;.\]

Due to the fact that $C_n + C_m \leq C_{m+n}$, the limit
\[ C_{\textrm{class}}(\Lambda)= \lim_{n \rightarrow \infty} \frac{C_n(\Lambda)}{n} \]
exists. Using Shannon's coding theorem, we see that $C$ is the least upper bound of the rate of information that can be transmitted with asymptotically vanishing error.

The HSW theorem \cite{HSW} gives an expression for this classical product state capacity of noisy memoryless quantum channels,
\[C_{\textrm{class}}(\Lambda)= \chi^* = \sup_{\pi,\rho} \chi(\Lambda),\]
where $\chi$ is the Holevo  $\chi$ quantity
\begin{align*}
&\chi(\{(\pi_i,\Lambda(\rho_i))\}) \\
 &= \Entropy(\sum_i \pi_i \Lambda(\rho_i)) - \sum_i \pi_i\Entropy(\Lambda(\rho_i)) \;.
\end{align*}
Due to the convexity of the von Neumann entropy, the supremum can in fact be taken over pure states $\rho_i$.

The memoryless channel restriction has recently been weakened to include, so called, \textit{forgetful} memory channels. For such channels, the classical product state capacity has been shown \cite{DKRW} to correspond to
\begin{equation}
\label{eq:capacity_forgetful}
C^*=\lim_{n\rightarrow \infty}\frac{C_{\textrm{class}}(\Lambda_n)}{n} \;,
\end{equation}
where $\Lambda_n$ is a channel representing the transmission of $n$ states, with the noise on subsequent transmissions is correlated. See \cite{DKRW} for details or Section \ref{sec:depolarizing} for an example.

\section{The Depolarizing Memory Channel}
\label{sec:depolarizing}

Treating information or noise sources as independent random variables is a successful but crude first approximation. To improve the modelling process and to achieve better performance in real world applications, the independence assumption needs to be removed. The first step in this direction is to introduce forgetful noise memory. A forgetful noise process is one which after sufficiently long time, `forgets' or is independent of previous noise. Thus, here the independence is pushed further away, allowing a space to study the effects of short-term memory. With the theoretical tools in place, it is instructive to study even very simple models to see the effects of memory on the classical capacity.

\subsection{Construction of the Channel}
\label{subsec:construction}

The forgetful channel is constructed by combining two memoryless single qubit depolarizing channels ($ \mathcal{E}_0$ and $ \mathcal{E}_1$), switching between them using a two-state Markov chain ($Q=(q_{ij}), \quad i,j \in \lbrace 0,1\rbrace$). Thus, $Q$ is the $2 \times 2$ Markovian channel selection matrix with $q_{ij}$ being the probability of switching from channel $i$ to channel $j$. Hence, $q_{ij} \geq 0$ and $q_{i0} + q_{i1} = 1$ for $i,j \in \lbrace 0,1 \rbrace$. It is forgetful, in the case when the Markov chain is \textit{aperiodic} and \textit{irreducible}.

The depolarizing channels can be written as: $ \mathcal{E}_i(\rho) = x_i^0 \rho + x_i^1 (\mathbf{1} - \rho)$. These single qubit channels can be thought of as probabilistically mixing the identity channel (with probability $x_i^0$) and `flip' channel (with probability $x_i^1=1-x_i^0$) acting on a single qubit density operator $\rho$. However this rewriting is only completely positive for $1/3 \leq x^0_i \leq 1$. 

The built-up channel $\Lambda_n$, corresponding to $n$ successive uses is
\begin{align*}
&\Lambda_n = \rho_1\otimes\ldots\otimes\rho_n \mapsto \\
&\sum_{i_1,\ldots,i_n} \gamma_{i_1} q_{i_1i_2} \ldots q_{i_{n-1}i_n}  \mathcal{E}_{i_1}(\rho_1)\otimes\ldots\otimes \mathcal{E}_{i_n}(\rho_n) \; .
\end{align*}
The sum is over all possible paths $(i_1,\ldots,i_n)\in \lbrace 0,1\rbrace^n$ and each term is a tensor product of the selected sub-channels weighted by the probability of occurrence ($\gamma_i$ is the initial probability of selection set to the stationary distribution of the Markov process: $Q^T \gamma=\gamma$). 


\subsection{Classical Capacity}
We calculate the capacity with this $n$-use form of the channel and regularize by taking the limit $n \rightarrow \infty$ as in \eref{eq:capacity_forgetful}. Since we are looking at the product state capacity, we choose
\begin{align*}
\rho_i = \Phi^{(n)}(\overline{l}) &= \Phi^{(n)}(l_1,\ldots,l_n) \\
& := \dop{l_1}\otimes\ldots\otimes\dop{l_n} \;,
\end{align*}
where the $l_i$ are arbitrary pure qubit states.

Applying the channel $\Lambda_n$, we get
\begin{multline*}\Lambda_n(\Phi^{(n)}(\overline{l})) = \sum_{i_1,\ldots,i_n} \gamma_{i_1} q_{i_1i_2}\ldots q_{i_{n-1}i_n} \\
(x_{i_1}^0 \dop{l_1 \oplus 0} + x_{i_1}^1\dop{l_1 \oplus 1})\otimes\ldots\\
\otimes(x_{i_n}^0\dop{l_n \oplus 0} + x_{i_n}^1\dop{l_n
\oplus 1}) \;,
\end{multline*}
where $(l_i \oplus 1)$ denotes the qubit state with a flipped Bloch vector with respect to $l_i=(l_i \oplus 0)$
\[ \dop{l_i \oplus 1} = \idty - \dop{l_i \oplus 0} \]

By expanding the product above we see that the eigenvalues of the output state are given by
\begin{equation}
\label{eq:eigenvalues}
\lambda_n(\overline{k}) = \sum_{i_1,\ldots,i_n} \gamma_{i_1}
q_{i_1i_2}\ldots q_{i_{n-1}i_n} x_{i_1}^{k_1}\ldots
x_{i_n}^{k_n} \;.
\end{equation}
Note that these eigenvalues are independent of the choice of the input state.

The channel output can now be written as
\[ \Lambda_n\left(\Phi^{(n)}(\overline{l})\right) = \sum_{\overline{k}} \lambda_n(\overline{k}) \Phi^{(n)} (\overline{l} + \overline{k}) \;. \]
Hence, if we calculate the first term in the Holevo $\chi$ quantity for $\pi$, the uniform distribution ($\pi_i=1/{2^n}$), and $\Phi_i$ going over all the $\rho^{(n)}(\overline{l})$, we see that
\begin{align*}
 \Phi_{\textrm{out}} &:= \sum_{\overline{l}} \frac{1}{2^n} \Lambda_n \left( \Phi^{(n)}(\overline{l}) \right)\\
&= \frac{1}{2^n} \sum_{\overline{k}} \lambda_n(\overline{k}) \sum_{\overline{l}} \Phi^{(n)} (\overline{l}+\overline{k}) \;.
\end{align*}
Since $\overline{l}$ goes over all possible combinations, so does $\overline{l}+\overline{k}$, so we can relabel them
\[ \Phi_{\textrm{out}} =  \frac{1}{2^n} \sum_{\overline{k}} \lambda_n(\overline{k}) \sum_{\overline{l}'} \Phi^{(n)} (\overline{l'}) \;. \]
Since the eigenvalues in \eref{eq:eigenvalues} sum to one, we see that $\Phi_{\textrm{out}}$ is the maximally mixed state
\[ \Phi_{\textrm{out}} =  \frac{1}{2^n} \sum_{\overline{l}'} \Phi^{(n)} (\overline{l'}) \;. \]
Thus, $S(\Phi_{\textrm{out}})$ is maximal and is equal to $\log_2(2^n)=n$.

The second term in the Holevo $\chi$ quantity is
\[ - \sum_i \pi_i S\left( \Lambda_n (\rho_i) \right) \;.\]
Since the eigenvalues $\lambda_n(\overline{k})$ of $\Lambda_n (\rho_i)$ do not depend on the choice of $\rho_i$, this term does not influence the maximization. Hence our choice of $\pi$ and $\rho$ maximizes the Holevo $\chi$ quantity.

Thus, the final expression for the regularized capacity \eref{eq:capacity_forgetful} is

\begin{equation}
\label{eq:capacity}
C^*=\lim_{n\rightarrow \infty}\frac{1}{n}C_{\textrm{class}}(\Lambda_n))= 1- \lim_{n\rightarrow \infty}\frac{1}{n} \Entropy(\Lambda_n(\rho))\; .
\end{equation}
 
If we were to calculate the output entropy using the eigenvalues in \eref{eq:eigenvalues}, the calculation would be exponentially long in $n$. Therefore, other techniques are needed. The way we approach the problem is by reformulating it as a hidden Markov process. The eigenvalues of the output state correspond to the probabilities of such a process.

A hidden Markov process can be defined as follows. If we have a translation-invariant measure $\nu$ with the Markov property on $L^{\mathbb{Z}}$, where $L$ is a finite set, then a hidden Markov measure can be constructed on $K^{\mathbb{Z}}$ through a function $\Phi : L \rightarrow K$, with the following local densities
\begin{equation}
\label{eq:hidden_markov}
\mu((\omega_m, \ldots, \omega_n)) = \sum_{\substack{\epsilon_m,\ldots,\epsilon_n\\ \Phi(\epsilon_m)=\omega_m \ldots \Phi(\epsilon_n)=\omega_n}} \nu((\epsilon_m,\ldots,\epsilon_n)) \;,
\end{equation}
where $\omega_m, \ldots, \omega_n \in K$ and $\epsilon_m,\ldots,\epsilon_n \in L$ .
For obvious reasons, these processes are also called functions of Markov processes.

\section{Algebraic Measures}
\label{sec:algebraic}
An algebraic measure, $\mu$, is a translational-invariant measure on a set $\lbrace 0, \ldots, q-1\rbrace^\mathbb{Z}$, with probabilities determined by matrices $E_a$ with positive entries, one for each of the $q$ states. The probability of a sequence is obtained by applying a positive linear functional $\sigma$ to a matrix product of the corresponding matrices of the states of the sequence: $\mu(i_1,\ldots,i_n)=\sigma(E_{i_1}\ldots E_{i_n})$. This matrix algebraic construction is the reason for the name \textit{Algebraic Measure}, studied in detail in Ref. \cite{FNS91}. As we shall see, the hidden Markov processes correspond to a set of algebraic measures with a specific positivity structure and remarkably, the converse holds too.

\subsection{Manifestly Positive Measures}
\label{sec:manifest}
In \cite{FNS91} it was shown that hidden Markov processes correspond to manifestly positive algebraic measures. The local densities of such a manifestly positive algebraic measure on an infinite chain $K^{\mathbb{Z}}$ of classical state spaces $K = \lbrace 0,\ldots,q-1 \rbrace$ are of the form
\[ \mu((\omega_1, \ldots , \omega_n)) = \bra \tau | E_{\omega_1} \ldots E_{\omega_n} \sigma \ket \;, \]
where $\omega_i \in K$, $\tau$ and $\sigma$ are vectors in $\mathbb{R}^{d}$ with non-negative elements (denoted $(\mathbb{R}^{d})^{+} $) and the $E_i$ are $d \times d$ real matrices with non-negative elements (denoted $M_d^{+} $).

As an example of these manifestly algebraic measures, let us look at a regular Markov chain $\mu((\omega_m, \ldots, \omega_n))$ on $\lbrace 0, \ldots, q-1\rbrace^{\mathbb{Z}} $. If we choose $\tau$, $\sigma$ and the $E_i$ as
\begin{align*}
&\sigma \in (\mathbb{R}^{d})^{+}: \;\; \sigma_a = 1 \text{ for } a \in K \;, \\
&\tau \in (\mathbb{R}^{d})^{+}: \;\;  \tau_a = \mu((a)) \text{ for } a \in K \;, \\
&E_a \in M_d^{+}: \;\; (E_a)_{b,c} = \delta_{a,b} \frac{\mu((b,c))}{\mu((b))} \text{ for } a,b,c \in K\,,
\end{align*}
one can check that $\bra \tau | E_{\omega_m} \ldots E_{\omega_n} \sigma \ket$ indeed gives the correct densities.

From this example it is easy to see that if we have a hidden Markov process on $L^{\mathbb{Z}}$ defined by a map $\Phi: K \rightarrow L$ and a Markov measure $\mu$ on $K$ with corresponding matrices $E_a$, the manifestly positive algebraic measure corresponding to the hidden Markov measure is given by the same vectors $\sigma$ and $\tau$ as before and the following matrices:
\begin{equation}
\label{eq:hidden_algebraic}
F_a \in M_d^{+}: F_a = \sum_{\epsilon, \Phi(\epsilon) = a} E_\epsilon \; \text{ for } a \in K \;.
\end{equation}

For a proof of the converse, which is namely, that every manifestly positive algebraic measure corresponds to a hidden Markov measure, we refer to \cite{FNS91}.

\subsection{Mean Entropy}
\label{sec:mean_entropy}
We will now briefly summarize how the algebraic measure approach allows for a simpler approach to finding the entropy density \cite{FNS91,DB57}.

The entropy of a state $\mu$ on $K^{\mathbb{Z}}$ restricted to a region $\Lambda$ is defined by
\[ S_{\Lambda} (\mu) = - \sum_{\omega_\Lambda \in K^{\Lambda}} \mu(\omega_\Lambda) \log \mu(\omega_\Lambda) \;. \]

$S_{\Lambda}$ can be shown to be bounded by $\#\Lambda \log q$, monotonically increasing in $\Lambda$ and strongly subadditive, that is
\[ S_{\Lambda_1 \cap \Lambda_2}(\mu) + S_{\Lambda_1 \cup \Lambda_2}(\mu) \leq S_{\Lambda_1}(\mu) + S_{\Lambda_2}(\mu) \;. \]

Using the strong subadditivity of the entropy and the translational invariance of the measure, one can show that \cite{RAMF,KAY}
\[ \Entropy(\mu)=\lim_{n\rightarrow \infty} \frac{\Entropy(\mu_n)}{n} = \lim_{n\rightarrow \infty} \Entropy(\mu_n) - \Entropy(\mu_{n-1}) \;.\]

We can then use this relation together with the expression for the local densities of the manifestly positive measures to reformulate the convergence of the mean entropy into a dynamical system of converging measures on the set of $d$-dimensional probability measures $\mathcal{B}_\sigma$ as
\[ \Entropy(\mu) = \lim_{n\rightarrow \infty} \sum_{a \in K} \int_{\mathcal{B}_\sigma} \phi_n (d\nu) h_a(\nu) \;, \]
where
\begin{align*}
\mu((\epsilon_0,\ldots,\epsilon_n)) &= \bra \tau | E_{\epsilon_0} \ldots E_{\epsilon_n} \sigma \ket \\
& \text{ with } \sigma, \tau \in (\mathbb{R}^d)^{+} \\
\mathcal{B}_\sigma &= \{\nu \in (\mathbb{R}^d)^{+} |\, \bra \nu | \sigma \ket = 1\} \\
h_a(\nu) &= - \bra \nu | E_a \sigma \ket \log \bra \nu | E_a \sigma \ket \\
\phi_n(d\nu) &= \sum_{\epsilon_0,\ldots,\epsilon_n \in K} \mu((\epsilon_0,\ldots,\epsilon_n)) \\
 & \hspace{2cm}\delta_{\frac{E_{\epsilon_n}^* \ldots E_{\epsilon_0}^*}{\mu((\epsilon_0,\ldots,\epsilon_n))}}(d\nu) \;.
\end{align*}

If we define the linear transformation $T_\mu$ on functions on $\mathcal{B}_\sigma: (T_\mu f)(\nu) = \sum_{a \in K} \bra \nu | E_a \sigma \ket f \left( \frac{E_a^* \nu}{\bra \nu | E_a \sigma \ket} \right)$, one can show that $\phi_n (f) = \phi_0(T_\mu^n f)$. $T_\mu$ is a contraction map, so a fixed point argument can be used to show that $\phi_n$ converges to a unique measure $\phi$ that is invariant under $T_\mu$
\[ \phi(T_\mu f) = \phi(f) \;. \]
This measure allows us then to calculate the mean entropy
\begin{equation}
\label{eq:entropy}
\Entropy(\mu) = \sum_{a \in L} \int_{\mathcal{B}} \phi(d\nu) h_a (\nu) \;.
\end{equation}

Our goal in the remaining part of the article is to translate the switching depolarizing channel into the setting of algebraic measures and to try and find the invariant measure that allows us to calculate the mean entropy.

\section{Algebraic Measure of the Channel}
\label{sec:our_algebraic}

 The relationship between the hidden Markov measure, say $\mu'$ on $K^{\mathbb{Z}}$, and the underlying Markov measure $\nu$ with the Markov property on $L^{\mathbb{Z}}$ is through a `tracing' function $\Phi : L \rightarrow K$, as is shown in \eref{eq:hidden_markov}.

The underlying Markov process for the overall quantum channel has a four state configuration space corresponding to channel selection and error occurrence: $K = \lbrace (0,0), (0,1), (1,0), (1,1) \rbrace$. The first index indicates which depolarizing channel has been chosen and the second indicates whether a bit flip occurred. The elements of the transition matrix, $E$, for this process are then given by
\begin{equation}
\label{eq:trans_elements}
(E)_{\{(ij)(i'j')\}} = q_{ii'}x_{i'}^{j'} \;,
\end{equation}
the probability of going from $(i,j)$ to $(i',j')$ is given by the switching probability $q_{ii'}$ from channel $i$ to $i'$, multiplied by the probability $x_{i'}^{j'}$ that channel $i'$ produces the error-occurrence $j'$.

The function that produces the correct hidden Markov process is then given by
\[ \Phi((i,j))=j \; .\]
This function reflects the fact that we are unaware of the choice of channel that has been made. The only effect that is visible from the outside is whether or not an input qubit has been flipped. Thus, $\Phi$ has to `trace out' the choice of channel. $\Phi$ maps into the two-state error configuration space containing `no flip' and `flip':  $L = \lbrace 0,1 \rbrace$ .

Using the fact that the matrices $E_{(i,j)}$ defining the algebraic measure of a Markov process (\secref{sec:manifest}, $a=(i,j)\in K$) have only one non-zero row and \eref{eq:hidden_algebraic}, we get the matrices $F_0$ and $F_1$ that define the algebraic measure corresponding to $\mu'$. The matrix corresponding to $0$, the first element of $L$ is given by
\begin{align*}
F_0&=\sum_{(i,k),\Phi'((i,k))=0} E_{(i,k)} = \sum_i E_{(i,0)}\\
&=
\begin{pmatrix}
q_{00}x_0^0 & q_{00}x_0^1 & q_{01}x_1^0 & q_{01}x_1^1\\
0&0&0&0\\
q_{10}x_0^0 & q_{10}x_0^1 & q_{11}x_1^0 & q_{11}x_1^1\\
0&0&0&0
\end{pmatrix}
\end{align*}
and similarly for 1, the second element of $L$.

The hidden Markov process then gives us almost the same probabilities as the eigenvalues in \eref{eq:eigenvalues}
\begin{align*}
	p((k_1,\ldots,k_n)) 	&= \bra \tau | F_{k_1} \ldots F_{k_n} \idty \ket \\
				&= \sum_{i_1,\ldots,i_n} \tau_{i_1,k_1}
q_{i_1i_2}\ldots q_{i_{n-1}i_n} x_{i_2}^{k_2}\ldots
x_{i_n}^{k_n} \;.
\end{align*}
Note that according to our discussion in Section \ref{sec:algebraic}, the vector $\tau$ is the stationary distribution of the full matrix $E$. Using \eref{eq:trans_elements}, one can see that the invariant distribution $\tau$ is in fact $\tau_{(i,k)}=\gamma_i x_i^k$, so the probabilities of the hidden Markov process coincide with the eigenvalues in \eref{eq:eigenvalues}.

Having constructed the correct algebraic measure, we can determine $T_\mu$ explicitly and use it to greatly simplify the corresponding invariant measure $\phi$.

The expression for $T_\mu$, as can be found in \cite{FNS91}, is
\[ (T_\mu f) (\hat{\nu}) = \sum_{a \in K} \bra \hat{\nu} | F_a 1 \ket f \left( \frac{F_a^* \hat{\nu}}{\bra \hat{\nu} | F_a 1 \ket} \right ) \;,\]
where $\hat{\nu}$ is any $4$-dimensional vector such that $\bra \hat{\nu} | 1 \ket = 1$ and $f$ is a continuous real-valued function on the set of such vectors. For the case of our hidden Markov measure, the form of this transformation can be greatly simplified. Due to the stochasticity of the matrix $E$, we have the following:
\[ F_0 |1\ket = \begin{pmatrix}
1\\
0\\
1\\
0
\end{pmatrix}
\text{   and   }
F_1 |1\ket = \begin{pmatrix}
0\\
1\\
0\\
1
\end{pmatrix} \;.\]
If we furthermore denote the four row vectors of $E$ by $\hat{\mu}_1$, $\hat{\mu}_2$, $\hat{\mu}_3$ and $\hat{\mu}_4$, we can write
\[ F_0^* \hat{\nu} = \nu_1 \hat{\mu}_1 + \nu_3 \hat{\mu}_3 \text{   and   } F_1^* \hat{\nu} = \nu_2 \hat{\mu}_2 + \nu_4 \hat{\mu}_4 \;.\]
On top of this, $\mu_1 = \mu_2$ and $\mu_3 = \mu_4$, so the total form of the transformation becomes
\begin{align*}
(T_\mu f)(\hat{\nu}) = &(\nu_1+\nu_3) f\left( \frac{\nu_1 \hat{\mu}_1 + \nu_3 \hat{\mu}_3}{\nu_1+\nu_3} \right)\\
&+ (\nu_2+\nu_4) f\left( \frac{\nu_2 \hat{\mu}_2 + \nu_4 \hat{\mu}_4}{\nu_2+\nu_4} \right) \;.
\end{align*}

From this form of the transformation, we can already greatly restrict the support of $\phi$. Our claim is that the support of $\phi$ is restricted to the set of convex combinations of $\mu_1$ and $\mu_3$
\[ \text{supp}(\phi) \subset \lbrace a \hat{\mu_1} + (1-a) \hat{\mu_3}\, |\, a \in [0,1]  \rbrace\;. \]

To show this, let's suppose that $\hat{\nu} \in \textrm{supp}(\phi)$ and $\hat{\nu} \not\in S := \lbrace a \hat{\mu_1} + (1-a) \hat{\mu_3}\, |\, a \in [0,1]  \rbrace $. Take $\zeta_{\hat{\nu}}$ a function on $\mathcal{B}_\sigma$ such that $\zeta_{\hat{\nu}}(\hat{s})=0$ for all $\hat{s} \in S$ and $\zeta_{\hat{\nu}} (\hat{\nu}) \neq 0$, then
\begin{align*}
0\neq&\phi(\zeta_{\hat{\nu}}) =\phi(T_\mu\zeta_{\hat{\nu}}) =\int \phi(d\nu)T_\mu(\zeta_{\hat{\nu}}(\nu))\\
=&\int \phi(d\nu)\Bigr[(\nu_1+\nu_3)\zeta_{\hat{\nu}}\Bigl(\frac{\nu_1\hat{\mu}_1+\nu_3\hat{\mu}_3}{\nu_1+\nu_3}\Bigr)  \\ &+ (\nu_2+\nu_4)\zeta_{\hat{\nu}}\Bigl(\frac{\nu_2\hat{\mu}_1+\nu_4\hat{\mu}_3}{\nu_2+\nu_4}\Bigr)\Bigr]\;.\\
\end{align*}
However, this integral is equal to zero, since the arguments to $\zeta_{\hat{\nu}}$ run over the set $S$. 

Therefore, we have for $f \in \mathcal{C}(\mathcal{B})$,
\begin{equation}
\label{eq:lambda_measure}
\phi(f) = \int_0^1 d\lambda(a) f(a \hat{\mu}_1 + (1-a)\hat{\mu}_3) \;,
\end{equation}
with $\lambda$ a measure on $[0,1]$.

Now let us look at $\phi$ acting on the transformed $f$:
\begin{align}
\phi(T_\mu f) =& \int  \phi(d\nu) \Bigl[
(\nu_1+\nu_3)f\Bigl(\frac{\nu_1\hat{\mu}_1+\nu_3\hat{\mu}_3}{\nu_1+\nu_3}\Bigr) \nonumber\\
 &+ (\nu_2+\nu_4)f\Bigl(\frac{\nu_2\hat{\mu}_1+\nu_4\hat{\mu}_3}{\nu_2+\nu_4}\Bigr)\Bigr] \nonumber \\
&= \int_0^1 d\lambda(a)\Bigl[(\hat{\mu}_{a,1} + \hat{\mu}_{a,3})f\Bigl(\frac{\hat{\mu}_{a,1}\hat{\mu}_1 + \hat{\mu}_{a,3}\hat{\mu}_{3}}{\hat{\mu}_{a,1} + \hat{\mu}_{a,3}}\Bigr) \nonumber \\
&+(\hat{\mu}_{a,2} + \hat{\mu}_{a,4})f\Bigl(\frac{\hat{\mu}_{a,2}\hat{\mu}_1 + \hat{\mu}_{a,4}\hat{\mu}_{3}}{\hat{\mu}_{a,2} + \hat{\mu}_{a,4}}\Bigr)\Bigr] \;, \label{eq:lambda_iterate}
\end{align}
where
\begin{align*}
\hat{\mu}_{a} = a \hat{\mu}_1 + (1-a) \hat{\mu}_3 \;.
\end{align*}

By invariance \secref{sec:mean_entropy}, we can equate \eref{eq:lambda_measure} and the above \eref{eq:lambda_iterate} to discover an invariance concerning $\lambda$. We thus arrive at the following symmetry of $\lambda$:
\[\lambda=T[\lambda] = a \mapsto c_1(a) \lambda[f_1(a)] + c_2(a) \lambda[f_2(a)]\;.\]

The two functions $f_1$ and $f_2$ are relatively simple shrink functions about two separate points in the domain $[0,1]$, that shrink the $[0,1]$ domain into two (possibly overlapping) sub-intervals of $[0,1]$.

We can turn this analytic symmetry into a cyclic definition or iterative procedure to generate $\lambda$ up to some approximation $\lambda_n$.

\[\lambda_{n+1} = T(\lambda_n) \;.\]

We still have not defined $\lambda_0$, but taking a look the iterative procedure, we see that there exist fixed points of the two shrink functions, call them $a_1$ and $a_2$,
\[a_1=f_1(a_1)\quad a_2=f_2(a_2)\qquad a_1,a_2\in [0,1] \;.\]

With this observation the idea is to begin the iteration procedure with two Dirac delta's at these fixed points,
\[\lambda_0(a) = \frac{1}{2} \delta(a-a_1) + \frac{1}{2} \delta(a-a_2) \;.\]
Note that $\int \lambda_0(a) da = 1$, as a measure should be. Since there is unique convergence then the initial weightings should not matter \cite{FNS91}.

To see that this is a good starting point and to get further insight into the support of $\lambda$, it can be seen that the support will grow, but most importantly, once a point is within the support of $\lambda_m$ it remains there for all $n \geq m$. So if the procedure is taken to infinity the support is fixed and countably infinite. Thus, we arrive at the following expression for the full support,
\begin{align*}
\textrm{supp}(\lambda) = \{&a\in[0,1]:\exists n \in \mathbf{N}, \exists k_i\in \{0,1\}\forall i \in [1,n]\\
&f_{k_{n}}\circ f_{k_{n-1}}\circ \ldots \circ f_{k_1}(a_1~\textrm{or}~a_2) = a  \} \;.
\end{align*}

We use this iterative procedure to generate $\lambda_n$ and then use it in \eref{eq:lambda_measure} to approximate the measure. The entropy in \eref{eq:entropy} can then be calculated and finally we use the entropy to calculate the capacity through \eref{eq:capacity}. It is the capacity and its dependence on memory that we are interested in.

\section{Results}
\label{sec:results}

In constructing our channel we defined certain parameters. It is useful to introduce a new set of suggestive parameters in terms of the old and also to reduce their number by making some assumptions. Firstly, we assume that the sub-channels switch symmetrically, that is, the probabilities of reuse are the same for both sub-channels. This makes the Markov matrix doubly stochastic and allows us to use its non-one eigenvalue as a useful characterizing parameter $s$. Thus, we set $q_{00} \rightarrow(1 + s)/2$ and $q_{10} \rightarrow(1 - s)/2$. The domain of $s$ is  $(-1,1)$, with $s=0$ corresponding to no noise correlations. Secondly, we parametrize the error probabilities by their average and difference: $x_0^0 \rightarrow a + d$, $x_1^0 \rightarrow a - d$.


The main result is that the capacity increases with stronger noise-correlations. This manifests itself in two ways. Firstly, if we make the switching more correlated ($s$ away from $0$) the capacity increases and secondly, if we increase the difference between the two sub-channels the capacity also increases. Similar results have been found for the quantum capacity of the dephasing channel with Markovian memory \cite{ABF}.

\begin{figure}
\centering

\psfrag{x}[tl][tr]{{\large$\boldsymbol{s}$}}
\psfrag{y}[bl][br]{{\large$\boldsymbol{a}$}}
\psfrag{z}[tr][br]{{\large$\boldsymbol{C^*}$}}

\psfrag{a1}[br][br]{{\small$\frac{1}{3}$}}
\psfrag{a2}[br][br]{{\small$\frac{2}{3}$}}
\psfrag{a3}[br][br]{$1$}

\psfrag{s1}[tr][tr]{$-1$}
\psfrag{s2}[tr][tr]{$0$}
\psfrag{s3}[tr][tr]{$1$}

\psfrag{c1}[cr][cr]{$0$}
\psfrag{c2}[cr][cr]{$0.5$}
\psfrag{c3}[cr][cr]{$1$}

\includegraphics[width=\columnwidth]{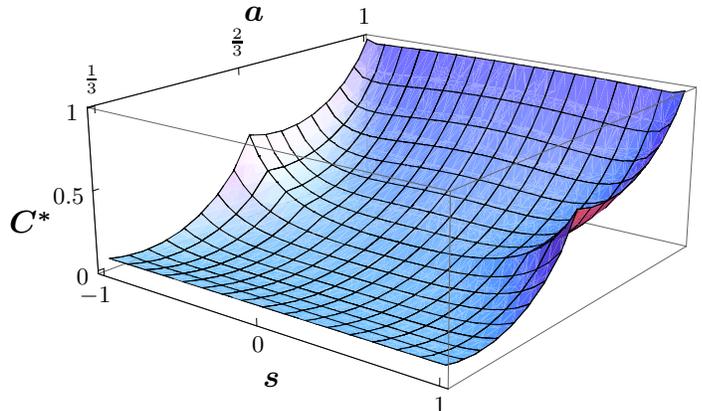}

\caption{Capacity for maximally different sub-channels increases with memory}
\label{fig:capacity3d}
\end{figure}

In Figure \ref{fig:capacity3d}, $d$ is set to the maximum possible value while keeping an average of $a$ ($d=\min[a-1/3,1-a]$). Remember that both $a-d$ and $a+d$ have to lie in the $[1/3,1]$ interval for the two sub-channels to be completely positive. The capacity is plotted against varying $a$ and $s$. We can see that the capacity increases as the noise-correlation ($s$) gets stronger. When $a=2/3$, $d$ attains its maximum ($1/3$) and the effect of increasing $s$ on the capacity is greatest. Another interesting observation is the case when the two sub-channels average to the maximally mixing channel ($a=1/2$, which ignoring memory, has zero capacity), taking into account memory effects there is a non-zero capacity.

\begin{figure}

\psfrag{T}[cc][cc]{}
\psfrag{xlabel}[cl][cl]{{\large$\boldsymbol{a}$}}
\psfrag{ylabel}[bc][tc]{{\large$\boldsymbol{C^*}$}}

\psfrag{b1}[cr][cr]{$0$}
\psfrag{b2}[cr][cr]{$0.08$}
\psfrag{b3}[cr][cr]{$0.54$}
\psfrag{b4}[cr][cr]{$1$}

\psfrag{x1}[tc][tc]{$0.5$}
\psfrag{x2}[tc][tc]{$0.75$}
\psfrag{x3}[tc][tc]{$1$}

\psfrag{LN}{{\tiny Low Noise Sub}}
\psfrag{AC}{{\tiny Avg Capacity}}
\psfrag{MC}{{\tiny With Memory}}
\psfrag{NM}{{\tiny Avg Channel}}
\psfrag{NS}{{\tiny Noisier Sub}}

\includegraphics[width=\columnwidth]{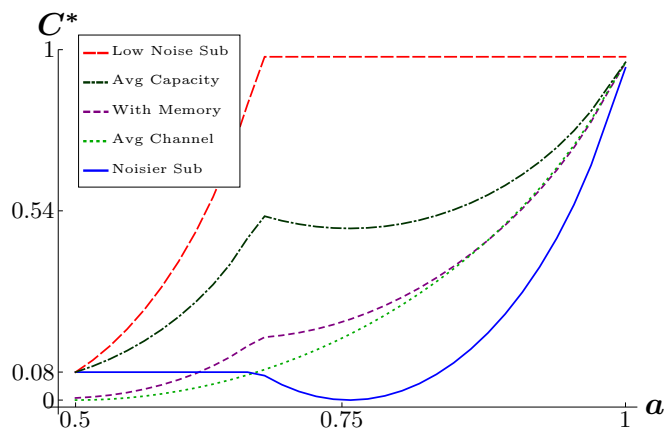}
\caption{Capacity versus the average no-error probability $a$}
\label{fig:capacity_vs_a}
\end{figure}

To better illustrate the last point and to further explore the relationship between the capacity of the memory channel and its sub-channels, we plot in Figure \ref{fig:capacity_vs_a}, slices of Figure \ref{fig:capacity3d} of fixed $s$ together with plots of the underlying sub-channel capacities.

Thus, in the `Avg Capacity' curve of Figure \ref{fig:capacity_vs_a}, we see the edge of Figure \ref{fig:capacity3d} (for fixed $s=1$, equivalently $s=-1$, not actually attained), which corresponds to the average of the capacities of the sub-channels. The sub-channels' separate capacities are plotted in curves labelled `Low Noise Sub' and `Noisier Sub'. They are chosen to have maximum allowed separation for each point as $a$ varies (and thus the artificial discontinuities). In a real world example, this separation parameter is fixed by the channel and the sub-channels and their capacities would not be accessible. The capacity of the average channel, labelled `Avg Channel', corresponds to a slice of fixed $s=0$ (the center of Figure \ref{fig:capacity3d}), since a no-memory/non-biased Markov walk factors into a tensor product of the average of the sub-channels, which is thus equivalent to just one depolarizing channel with the average error probability. The curve, `With Memory', is a smooth intermediary between the `Avg Channel' and `Avg Capacity' and is an example slice of Figure \ref{fig:capacity3d} for $0<s=\frac{2}{3}<1$, which illustrates how taking memory into account improves the capacity. Of course, again, in a real world example this parameter is specified by the channel. The smooth transformation is not straightforward nor linear, which can be seen in the way Figure \ref{fig:capacity3d} curves for varying $s$.

\begin{figure}


\psfrag{T}[cc][cc]{}
\psfrag{xlabel}[cc][cl]{\large$\boldsymbol{s}$}
\psfrag{ylabel}[cc][cc]{\large$\boldsymbol{C^*}$}
\psfrag{y2}[cr][cr]{$0.25$}
\psfrag{b1}[cr][cr]{$0.08$}
\psfrag{b2}[cr][cr]{$0.008$}
\psfrag{b3}[cr][cr]{$0.49$}
\psfrag{x2}[cc][cc]{$0.5$}
\psfrag{x3}[cc][cc]{$1$}

\psfrag{IT}{{\tiny \underline{\textbf{Iterations}}:}}
\psfrag{MF}{{\tiny Markov Full}}
\psfrag{HL}{{\tiny \underline{\textbf{Horiz. Lines:}}}}

\psfrag{I1}{{\tiny 1}}
\psfrag{I2}{{\tiny 2}}
\psfrag{I3}{{\tiny 3}}
\psfrag{I4}{{\tiny 4}}
\psfrag{I5}{{\tiny 5}}

\psfrag{LN}{{\tiny Low Noise Sub}}
\psfrag{AC}{{\tiny Avg Capacity}}
\psfrag{NM}{{\tiny Avg Channel}}
\psfrag{NS}{{\tiny Noisier Sub}}

\includegraphics[width=\columnwidth]{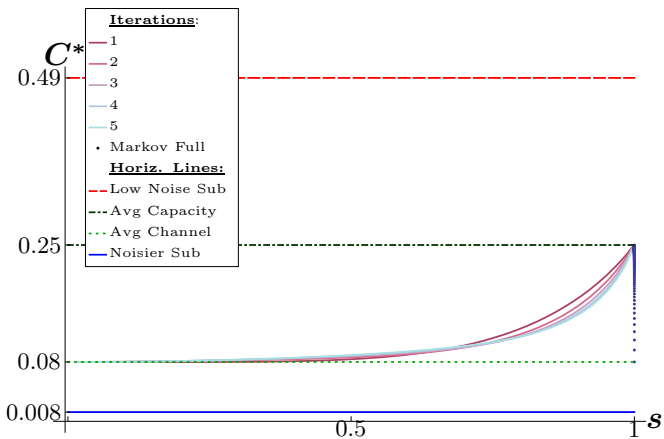}
\caption{Capacity versus the memory parameter $s$ using many iterations and including full Markov calculation}
\label{fig:capacity_vs_s}
\end{figure}

To see the last point more clearly and also to indicate the convergence of the iteration procedure we next plot a slice of Figure \ref{fig:capacity3d} for fixed $a$. In Figure \ref{fig:capacity_vs_s} we plot the regularized capacity against $s$ with the following fixed parameters: $a=\frac{2}{3}$, $d=\frac{1}{3}$.

We can see that the capacity increases as the noise-correlation gets stronger.
The blue dots are calculated using a simplified ($s=1$) full Markov walk calculation ($1000$ steps) which doesn't suffer from the usual exponential blow-up. The horizontal green line is the output entropy for $s=0$, which is corresponds to no correlations and is equivalent to having only one depolarizing channel.

\subsection{Non-Forgetful Limit}
To complete the discussion concerning correlations we need to look at the two extreme cases: $s=1$, corresponding to the case where a sub-channel is selected and used for every channel use afterwards, and $s=-1$, corresponding to the case where the choice of sub-channel is flipped with every channel use. Therefore, in constructing the overall channel and taking into account the initial random channel selection, we just have the mixing of two $n$-use channels. Specifically, in the $s=1$ case, we have the mixing of the two $n$-fold tensor products of the two sub-channels separately, and in the $s=-1$ case, we have the mixing of two $n$-use channels where each deterministically alternates between the sub-channels but starting with a different sub-channel.

Both these extreme cases are non-forgetful since the initial sub-channel selection (the initial noise) is `remembered' and the forgetful Holevo capacity theorem no longer applies (the Markov selection matrix is periodic in the $s=-1$ case and reducible in the $s=1$ case). While our forgetful channel approach breaks down there are alternate theoretical frameworks that do actually capture these extreme cases. For $s=-1$ the capacity can be calculated using \cite{DD07} and agrees with the limit of the forgetful approach, the capacity is the average capacity of the two sub-channels separately. However, for $s=1$ case there is a discontinuity and the capacity suddenly drops to the minimum capacity of the sub-channels \cite{DD06}.

The intuition is that in the $s=-1$ case, the deterministic flip can be used to determine `on-the-fly' which sub-channel is being used and then it is the same as using the two channels separately each half the time, so the capacity must be the average capacity. For the $s=1$ case once you have the poorer channel you are stuck with it forever and so because of the mixture you can only guarantee the lower capacity.

\section{Conclusion}
We have constructed a simple forgetful noise-memory quantum channel. The noise-correlation is a function of the underlying hidden Markov process. This setup allowed us to construct a corresponding algebraic measure. We used the measure in an algebraic asymptotic entropy expression. Without this, the entropy would be very difficult to compute, involving exponentially many paths in configuration space.

We studied the effects that the noise correlations had on the classical capacity and discovered that the capacity increases with stronger correlations. This is sensible because the correlations can be used to combat the noise when coding information. We have arrived at the understanding that stronger correlations increases the capacity from that of the average channel to the average capacity of the sub-channels with very interesting limiting behaviour.

Further work includes using other approximation techniques, arriving at a full analytic expression of the capacity and looking at other similarly constructed channels. We are also confident and hopeful that the hidden Markov technique could be successfully employed in other contexts.

\begin{acknowledgements}
We would like to acknowledge N. Datta and T. Dorlas for the idea of the channel construction and valuable assistance. This work is based upon research supported by the South African Research Chair Initiative of the Department of Science and Technology and National Research Foundation.
\end{acknowledgements}

\end{document}